\documentstyle[epsf,epsfig,rotate]{mn}

\begin{document}
\title{Dynamical relaxation and the orbits of low--mass extrasolar
planets}

\author [C. Terquem \& J.C.B. Papaloizou] {
        Caroline Terquem
       \\ Institut d'Astrophysique de Paris, 98~bis Boulevard Arago,
       75014 Paris, France -- terquem@iap.fr \\ Universit\'e Denis
       Diderot--Paris VII, 2 Place Jussieu, 75251 Paris Cedex 5,
       France \and John C. B. Papaloizou \\
       Astronomy Unit, Queen Mary, University of London,  Mile End Rd,
       London E1~4NS, UK -- j.c.b.papaloizou@qmul.ac.uk}

\date{Accepted.
      Received;
      in original form }

\pubyear{}


\maketitle

\begin{abstract}

We consider the evolution of a system containing a population of
massive planets formed rapidly through a fragmentation process
occurring on a scale on the order of 100~au and a lower mass planet
that assembles in a disc on a much longer timescale.  During the
formation phase, the inner planet is kept on a circular orbit due to
tidal interaction with the disc, while the outer planets undergo
dynamical relaxation.  Interaction with the massive planets left in
the system after the inner planet forms may increase the eccentricity
of the inner orbit to high values, producing systems similar to those
observed.

\end{abstract}


\section{Introduction} \label{intro}

The extrasolar planets discovered so far have masses, semi--major axes
and eccentricities in the range 0.16--11~Jupiter masses ($M_{\rm J}$),
0.038--4.47~au and 0--0.93, respectively.  In general,  high
eccentricities are very difficult to explain in the context of a
model where an isolated planet forms in a disc, as the
disc--protoplanet interaction leads to strong eccentricity damping
(Nelson et al. 2000).

Papaloizou \& Terquem (2001, hereafter PT01) investigated a scenario
in which a population of planetary mass objects formed rapidly through
a fragmentation process occuring in a thick disc or protostellar
envelope on a scale of 100~au.  Such a system then underwent dynamical
relaxation on a timescale of hundreds of orbits which resulted in
ejection of most of the objects.  It was found that the
characteristics of massive eccentric extrasolar planets and the
massive 'hot Jupiter' observed so far might be accounted for by such a
model.  However, planets with masses smaller than a few~$M_{\rm J}$
are probably too small to have formed through a gravitational
instability or fragmentation process (Masunaga \& Inutsuka 1999, Boss
2000) and are thus more likely to have grown through the core
accretion mechanism in a protoplanetary disc (Mizuno 1980).

Here, we investigate a scenario in which we have a planet accumulated
in a disc (PAID) on a timescale of $10^6$~years, being the observed
lifetime of protostellar discs (Haisch, Lada \& Lada 2001), together
with a population of massive outer planets already formed through
fragmentation processes on a much shorter timescale.  During its
formation phase, the PAID is kept in a circular orbit by tidal
interaction with the disc while the system of outer planets undergoes
dynamical relaxation.  After disc dispersal occurring at
$t=10^6$~years, the eccentricity of the PAID can be pumped up to high
values by interaction with the remaining bound massive planets.

In order to focus on the interaction mechanisms, we analyse the motion
of a planet under a distant perturber in a highly eccentric orbit in
\S~\ref{sec:secular}.  In \S~\ref{sec:results} we present numerical
simulations of the evolution of a system containing outer massive
planets and an inner PAID interacting with each other.  Finally, in
\S~\ref{sec:disc} we discuss our results.


\section{Secular perturbation theory and long term cycles}\label{secu}

\label{sec:secular}

We consider the evolution of the orbit of a PAID around a central star
due to gravitational interaction with a massive long period perturber
on a highly eccentric orbit that may have resulted from a prior
relaxation process.  We illustrate three possible effects.  The first
is the appearance of long period cycles in which the inner planet may
attain high eccentricity.  The second, occuring for stronger
perturbations, is a sequence of dynamical interactions leading to the
collision of the inner planet with the central star.  The third,
occurring for weakly bound perturbers, is an interaction leading to
the gradual ejection of the the outer planet to large radii, where
additional perturbations from external objects could cause it to
become unbound, leaving the inner planet on a highly eccentric orbit.

To discuss these three scenarios, we denote the osculating semi--major
axes of the inner planet and the perturber by $a$ and $a_p$,
respectively. The corresponding eccentricities are $e$ and $e_p$ and
the masses are $m$ and $m_p$, respectively. The central mass is
$M_\star.$


When the pericentre distance of the perturber, $d_p=a_p (1-e_p)$,
significantly exceeds $a,$ the large difference in the orbital periods
reduces the significance of short term variations such that the
interaction can be discussed using secular perturbation theory, in
which one considers evolution of the time averaged orbits.

Consider first the case when the orbits are coplanar.  Then analytic
treatment is possible in the limit $m \ll m_p$.  We denote $(r,\phi)$
and $(r_p,\phi_p)$ the cylindrical coordinates of the inner and outer
mass, respectively, in a frame centered on the star.  The perturbing
potential energy or Hamiltonian of the inner planet, including the
indirect term (which accounts for the acceleration of the origin of
the coordinate system), is:

\begin{equation}
H= -{Gm m_p\over \sqrt{r^2 +r_p^2 -2 r r_p \mu}} +{G m m_p r \mu \over
r_p^2 },
\end{equation}

\noindent where $\mu =\cos(\phi - \phi_p).$ For $r/r_p \ll 1,$ we
expand $H$ in spherical harmonics retaining terms up to third order in
$r/r_p$:

\begin{equation}
H= -{Gm m_p \over r_p} \left[ \left( {r \over r_p} \right)^2 P_2
\left( \mu \right) + \left( {r \over r_p} \right)^3 P_3 \left( \mu
\right) \right] ,
\end{equation} 

\noindent 
where $P_n$ is the Legendre polynomial of order $n$.  We have omitted
the term $-Gmm_p/r_p$ as it does not depend on the coordinates of the
inner planet ($H$ is defined within the addition of a constant).  On
performing time averages over both orbits (see, e.g., Roy 1978), one
obtains correct to second order in $e$ but with no restriction on
$e_p$:
 
\begin{equation}
H= - {G m m_p a^2 \over \left( 1 - e_p^2 \right)^{3/2} a_p^3} \left[
\frac{1}{4} +\frac{3 e^2}{8} - {15 e e_p a \cos \varpi \over 16 a_p
\left( 1-e_p^2 \right)} \right].
\label{Haver}
\end{equation}

\noindent Here $\varpi$ is the angle between the apsidal lines of the
two orbits.  For $m \ll m_p$, we can consider that the orbit of the
outer planet is fixed, i.e. that $a_p$ and $e_p$ are constant.  The
secular evolution is characterized by a constant $a$ and oscillations
of $e$ (see, e.g., Murray \& Dermott 1999).  Since $de/dt \propto
\partial H / \partial \varpi$ (Lagrange's equation), $e$ is maximum
for $\varpi =0$.  During the evolution $H(e,\varpi)$ is constant.
Since $e$ passes through zero (the inner orbit has no initial
eccentricity), the maximum value of $e$, $e_{\rm max}$, can then be
calculated by writing $H(0, \varpi)=H(e_{\rm max},0)$.
This gives $e_{\rm max}= 5 a e_p/ [ 2 a_p (1-e_p^2) ]$.  For
illustrative example we take $e_p =0.9$ and $a/a_p=0.01$, so that
$d_p/a = 10$.  Then $e_{\rm max}= 0.118$ which is significant and
independent of $m_p$.

Below we consider the case $a=0.5$~au, $a_p=50$~au, $m=0.3$~M$_{\rm
J}$ and $m_p =7$~M$_{\rm J}$.  It turns out that for such small $a$,
relativistic effects have to be taken into account. This can be done
following the procedure given in Lin et al. (2000), according to which
the central potential is modified such that:

\begin{equation}                             
{-GM_\star \over r} \rightarrow {-GM_*\over r}\left(1+{3GM_*\over r
c^2}\right).
\label{relat}
\end{equation}

\noindent This amounts to adding the term $-3 (GM_\star)^2 m ( 1+
e^2/2 ) / (a^2 c^2)$ on the right hand side of equation~(\ref{Haver}).
From this one readily finds that $e_{\rm max}$ is reduced by a factor

\noindent 
$1+4[GM_\star/(ac^2)](M_\star / m_p) (a_p/a)^3 (1-e_p^2)^{3/2}$.  For
the above parameters and $M_\star = 1$~M$_{\odot}$, this factor is
1.93, giving $ e_{\rm max}=0.06$.

\begin{figure}
\centerline{
\epsfig{file=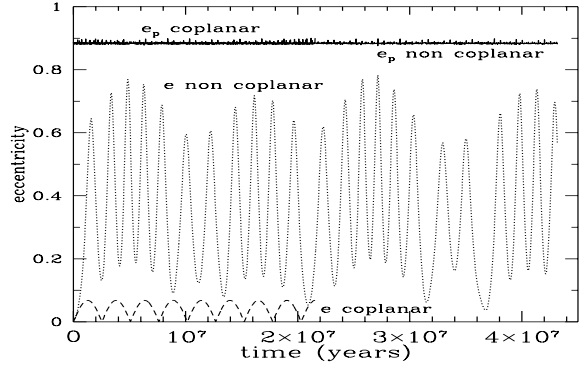,height=5.cm,width=8.cm} }
\caption[]{ Eccentricities versus time for $a/a_p =0.01,$ $m_p=
7$~M$_{\rm J}$ and $m=0.3$~M$_{\rm J}$, in the coplanar case (inner
planet, {\em short--dahed line}; outer planet, {\em solid line}) and
in the non coplanar case (inner planet, {\em dotted line}; outer
planet, {\em solid line}), where there is much larger variations.  In
both cases the eccentricity of the outermost massive planet remains
near to $0.9$.}
\label{fig1}
\end{figure}

We have calculated the orbital evolution numerically using the
Bullirsh--Stoer method (e.g., Press et al. 1993).  The outer planet
was started at pericentre and the inner one on a circular orbit at
random phase.  Of course, in contrast to the above analysis, both
orbits were allowed to vary.  Figure~\ref{fig1} shows the evolution of
the orbital eccentricities as a function of time.  The results are in
good agreement with the analysis, exhibiting eccentricity cyles with
amplitide $e_{\rm max} \sim 0.06$ as expected.  This is much lower
than typical values observed for extrasolar planets.  However, much
higher eccentricities can be obtained through the Kozai mechanism if
the orbits are mutually inclined (e.g. Lin et al. 2000 and references
therein).  Analysis for an outer circular orbit suggest high
eccentricities may be obtained for mutual inclinations exceeding
$\arccos \sqrt{3/5} \sim 40^{\circ}$.  We have also followed the
orbital evolution of the system when the orbits are initially mutually
inclined at an angle of $60^{\circ}$ and the results are also shown in
figure~\ref{fig1}.  In this case, large amplitude cyles with $e$
attaining values up to $0.7$ are obtained.  This is typical for runs
of this kind.  Cases with $a/a_p =0.01$ and $e_p =0.9$ lead to long
term secular variations that can provide high eccentricities of the
inner orbit.

\begin{figure}
\centerline{
\epsfig{file=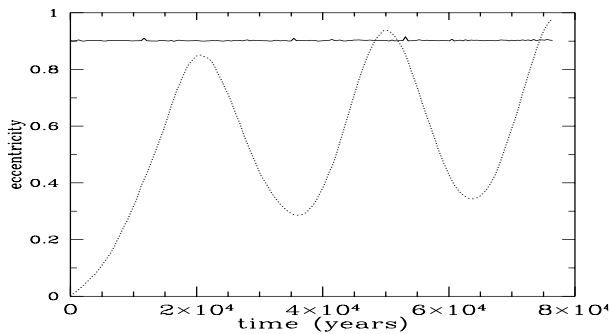, height=4.5cm,width=8.cm} }
\caption[]{ Eccentricities versus time for $a/a_p =0.04,$ $m_p=
7$~M$_{\rm J}$, $m=0.3$~M$_{\rm J}$ and initual mutual orbital
inclination of $60^{\circ}$.  The eccentricity of the inner planet
({\em dotted line}) approaches unity after about $t=7 \times 10^4$~y.
Subsequently, a collision with the central star occurs.  The
eccentricity of the outermost massive planet ({\em solid line})
remains near to $0.9$.}
\label{fig2}
\end{figure}

When $a/d_p$ is larger, the ensuing interaction with the inner planet
is stronger and can lead to very high eccentricities.  In such cases,
there can be a close approach or collision with the central star.  An
example of a run of this kind is illustrated in figure~\ref{fig2}.  In
this case the initial mutual orbital inclination, the initial values
of $a$ and $e_p$ and planet masses were as above, but the initial
value of $a_p$ was reduced by a factor of four.  The outer planet was
started at apocentre and the inner one on a circular orbit at random
phase.  This case led to very strong interactions at pericentre which
resulted in the eccentricity of the inner planet approaching unity and
a collision with the central star after about $t =7 \times 10^4$~y.
This situation corresponds to a more strongly bound outer planet than
previously.

\begin{figure}
\centerline{
\epsfig{file=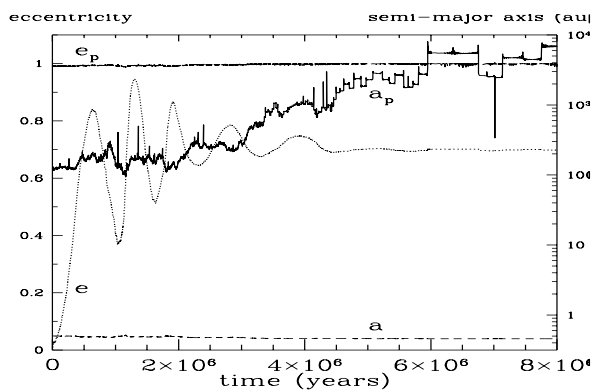, height=5.cm,width=8.cm} }
\caption[]{ Eccentricities (inner planet, {\em dotted line}; outer
planet, {\em long--dashed line}) and semi--major axes (inner planet,
{\em short--dashed line}; outer planet, {\em solid line}) versus time
in the case of a weakly bound outer planet.  Here $a/a_p =0.004,$
$m_p= 7$~M$_{\rm J}$, $m=0.3$~M$_{\rm J}$ and the initual mutual
orbital inclination is $60^{\circ}$.  In this case, $a$ decreases
slightly as energy transfers to the outer planet cause it to reach
radii exceeding $10^4$~au.  At this stage, $e$ reaches $\sim 0.7$,
whereas $e_p$ is always near to unity.}
\label{fig3}
\end{figure}

When more weakly bound outer planets are considered, the interactions
can also lead to high eccentricities of the inner planet.  As an
example we consider a case with $e_p =0.99$ and $a_p= 252 a$, other
parameters being as above.  The evolution is illustrated in
figure~\ref{fig3}.  Here, $d_p/a=2.52$, which is similar than in the
previous case.  However, here, the outer planet is very weakly bound.
Therefore, as a result of significant energy changes occurring at
pericentre passage in a chaotic manner, it eventually reaches radii
exceeding $10^4$~au, where it could become unbound if additional
perturbations from external objects occured.  The inner planet is left
on a highly eccentric orbit with $e \sim 0.7$.

\section{$N$ planets simulations}

\label{sec:results}

We consider a system of $N$ outer planets each having a mass
$m_p=7$--8~M$_{\rm J}$, one inner planet with a mass $m$, and a
primary star with a mass $M_\ast=1$~M$_\odot$, moving under their
gravitational attraction.  We take $m=0.3$~M$_{\rm J}$ (Saturn mass)
as, among the observed systems, this is the most extreme case of a
planet forming in a disc.  We suppose the outer massive planets have
formed rapidly through fragmentation on a scale of 100~au, while the
PAID forms in the disc surrounding the star on a much longer timescale.

\subsection{Initial conditions}

At $t=0$, we place $5 \le N \le 20$ massive planets taken at random
from a mass distribution corresponding to a uniform density spherical
shell with outer radius of $100$~au and inner radius of $10$~au.  The
planets, the orbits of which are not coplanar, are given the local circular
velocity in the azimuthal direction.  As shown by PT01,  adopting different
initial conditions, such as a  mass distribution in the form
of a flattened disc,
leads to the same qualitative evolution  as long as three dimensional
relaxation occurs , so our
results do not depend on these initial conditions.  The system is then
allowed to evolve under the influence of the gravitational attraction
of the central star and the different planets (the potential of the
star has been modified according to eq.~[\ref{relat}]).  We also
include the tidal interaction between the star (which radius is taken
to be 1~R$_\odot$) and a closely approaching planet (see PT01 for
details).  The approximation used is valid only when the planet
approaches the star on a highly eccentric orbit, which is the case
when tidal interactions first occur in the systems we consider here.
The system undergoes dynamical relaxation as described by PT01.

We assume a PAID forms on a timescale of $10^6$~y.  During the
formation process, any eccentricity that would be excited by the outer
massive planets is damped by the tidal interaction with the disc. We
therefore built up the planet progressively and constrained its orbit
to remain circular during this phase, with an orbital radius $a$ in
the range 0.3--10~au.  Depending on $a$, the planet may have either
formed at this location or migrated there as a result of tidal
interactions with the disc.

\subsection{Results}

We have run about 30 cases with the initial conditions described
above.  As found in PT01, most of the dynamical relaxation of the
system of outer planets is over after a few hundred orbits, i.e. a few
$10^5$~y.  This relaxation results in most of the planets being
ejected, while a few (usually between 0 and 3) of them end up on
highly, sometimes close, eccentric and  inclined  orbits
around the star.  These objects may still interact with each other
after most of the relaxation has occured.  As a result, or directly
because of the main relaxation, they may end up close enough to the
star to perturb significantly the PAID after it forms.

When there are still a few outer massive planets left at $t=10^6$~y,
if they make frequent incursions into the inner parts of the system,
the eccentricity of the PAID is pumped up to high values and the object
eventually hits the star (e.g. the calculation illustrated in
figure~\ref{fig2}).

\begin{figure}
\centerline{
\epsfig{file=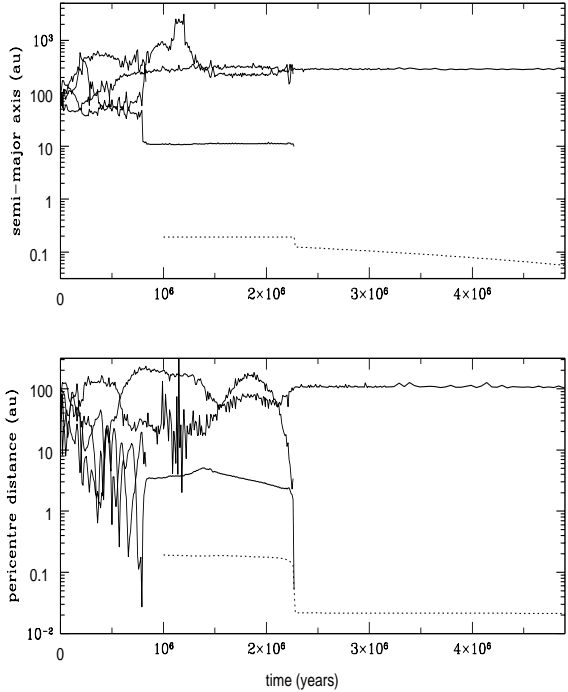,height=10.cm,width=8.cm} }
\caption[]{ Semi-major axes (in au, {\em upper plot}) and pericentre
distances (in au, {\em lower plot}) of the $N=4$ outer planets ({\em
solid lines}) and the PAID ({\em dotted lines}) versus time (in years).
Here $m_p=8$~M$_{\rm J}$ and $a=0.2$~au.  In this and other similar
figures, a line terminates just prior to the escape of a planet.  Due
to interactions with the outer planets, the orbit of the PAID acquires
a high eccentricity, and subsequently begins to be circularized by the
central star.  During this process, the semi--major axis decreases
whereas the pericentre distance stays constant equal to 0.02~au.  }
\label{fig4a}
\end{figure}

In some cases, the PAID may approach the star closely enough for tidal
effects to become important while a collision is avoided.  This is
illustrated in figure~\ref{fig4a}, which shows the evolution of the
semi--major axes and the pericentre distances of the planets.  Here
$m_p=8$~M$_{\rm J}$, $N=4$ and $a=0.2$~au.  The interaction with the
outer planets left in the system at $t=10^6$~y increases the
eccentricity of the PAID up to high values.  It approaches the star
closely enough for tidal circularization to start to occur.  This
causes the semi--major axis to decrease while maintaining the
pericentre distance almost constant equal to 0.02~au.  At this stage
the inclination between the orbits of the two planets is  close to
90$^\circ$.

\begin{figure}
\centerline{
\epsfig{file=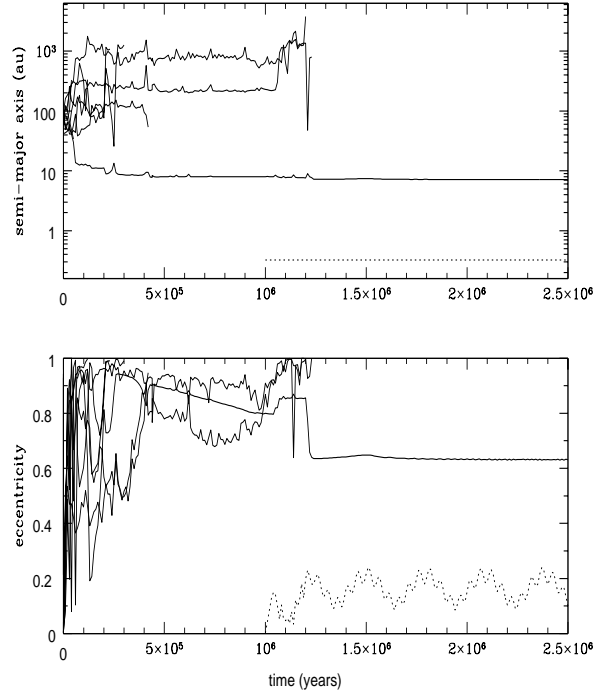,height=10.cm,width=8.cm} }
\caption[]{ Semi-major axes (in au, {\em upper plot}) and
eccentricities ({\em lower plot}) of the $N=9$ outer planets ({\em
solid lines}) and the PAID ({\em dotted lines}) versus time (in years).
Here $m_p=8$~M$_{\rm J}$ and $a=0.3$~au.  The PAID acquires an
eccentricity which varies in a cycle between 0.1 and 0.24 whereas its
semi--major axis stays constant equal to the initial value.  }
\label{fig4}
\end{figure}

If they are at most a few incursions down to small radii of the
massive planets left in the system at $t=10^6$~y, the PAID may
remain on an orbit with at most a moderate eccentricity and a constant
semi--major axis more or less equal to its initial value.  This case
is illustrated in figure~\ref{fig4}, which shows the semi--major axes
and the eccentricities of the different planets as a function of time.
Here $m_p=8$~M$_{\rm J}$, $N=9$ and $a=0.3$~au.  At $t=10^6$~years,
there are still two massive outer planets with high semi--major axes
and eccentricities interacting with each other and one massive planet
on an eccentric orbit with a semi--major axis of about 10~au.  Close
approaches of the two outermost planets (which eventually are ejected)
pump the eccentricity of the PAID up.  This eccentricity subsequently
goes into a cycle where it varies between 0.1 and 0.24 due to the
secular interaction with the third outer object.  This is the same
type of process as illustrated in figure~\ref{fig1}.  The semi--major
axis of the PAID stays constant equal to its intial value $a=0.3$~au
and the inclination between the orbits of the two planets
oscillates between 0 and 30$^\circ$.

\begin{figure}
\centerline{
\epsfig{file=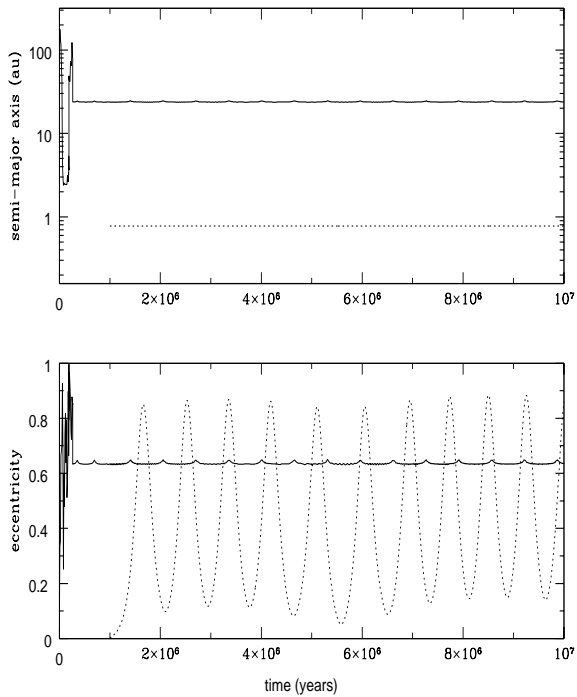,height=10.cm,width=8.cm} }
\caption[]{ Same as figure~\ref{fig4}, but for $N=29$, $m_p=7$~M$_{\rm
J}$ and $a=0.77$~au.  Among the massive planets, only the one which
was not ejected after about $3 \times 10^5$~years is represented.  Due
to the secular interaction with the massive planet, the orbit of the
PAID acquires an eccentricity which varies in a cycle between 0.05 and
0.88, whereas its semi--major axis stays constant equal to the initial
value. }
\label{fig5}
\end{figure}

When the few outer planets left from the relaxation process make at
most a few incursions in the inner system, the eccentricity of the PAID
may be increased up to high values by the secular interaction with one
of the massive objects with appropriate orbital parameters, but
without the process resulting in a collision with the star.  The
eccentricity then varies in a cycle as described in
\S~\ref{sec:secular}.  This also happens when there is only one outer
planet left at $t=10^6$~y, provided it is close enough to perturb
the PAID.  This case
is illustrated in figure~\ref{fig5}.  Here $m_p=7$~M$_{\rm J}$, $N=29$
and $a=0.77$~au.  Among the massive planets, we have represented the
only one which was not ejected after about $3 \times 10^5$~y.  We note
that the pericentre distance of the outer planet is about 9~au.  These
results are then in agreement with those of \S~\ref{sec:secular},
which showed similar behaviour of the eccentricity for such a ratio of
outer pericentre distance to inner semi--major axis  when the initial
inclination was 60$^\circ$ ( see figure \ref{fig1}).  
Here the inclination between the orbits 
of the two planets oscillates between
0 and 60$^\circ$. 

A complete survey of all the runs we have performed indicates that due
to the randomization of the inclinations of the orbits by dynamical
relaxation in three dimensions high inclinations always occur and an
eccentric PAID is produced provided the semi--major axis of its orbit
is in the appropriate range, that is not less than about 10 times
smaller than the pericentre distance of the innermost outer giant
planet and there need be only one of these.  If this condition is
satisfied, efficient eccentricity pumping occurs because of the
inclination of the orbit of the PAID with respect to that of the outer
planet.  An assessment of the probability for the semi--major axis of
the PAID to be in the appropriate range depends on as yet poorly
understood mechanisms of planet formation and migration in the disc in
the presence of perturbation by an outer giant planet.  More
information about this will be obtained with increasing planet
detections.  We have assumed that the disc is removed after
$10^6$~y. However, similar results are obtained if this occurs at a
later time as long as the PAID orbit is circularized when the disc is
present. This is the case as long as the disc is removed after the
major relaxation (see also \S~\ref{sec:secular}) .

\section{Summary and Discussion} 
\label{sec:disc}

The analysis and results presented above show that the eccentricity of
an inner PAID may be increased up to high values by secular interaction
with a more distant massive planet on an eccentric orbit.  The
mechanism is efficient when the ratio of outer pericentre distance to
inner semi--major axis is about 10 and the orbits of the planets are
significantly inclined.

Such a situation arises when the dynamical relaxation of a system of
outer massive planets takes place on a timescale shorter than the
timescale for assembling the PAID in a disc. After disc dispersal, as
was shown by PT01, the PAID would be expected to mostly interact with
one more massive object orbiting the star with semi-major axis on the
order of tens of au.  If the relaxation is not over by the time the
PAID forms, close approaches with the massive outer planets tend to
result in the object colliding with the star (see figure \ref{fig2}).

Since giant planet formation occurs on a timescale not longer than the
disc lifetime of $\sim 10^6$~y (Haisch et al. 2001), assuming
formation of massive planets occurs on a scale of 100~au, an eccentric
PAID at less than 1~au from the central star is more likely to be
produced when the number of outer planets in the system is $N \sim
15$, although $N=5$ may also produce such a system.

To illustrate the operation of the processes described here, we
consider observed planets with masses in the Saturn mass range.  Two
of them (HD~16141 and HD~6434) have orbital eccentricities of 0.28 and
0.3, respectively.  These modest eccentricity systems could be similar
to that represented in figure~\ref{fig4}.  Two others (HD~83443 and
HD~108147) have orbital eccentricities of 0.42 and 0.58 respectively.
These larger eccentricity systems systems could be similar to that
shown in figure~\ref{fig5}.  In these cases the eccentricities are
driven by secular interactions with a more closely approaching outer
planet.  Note however that the probability of getting a very high
eccentricity on a close orbit is reduced because the chance of hitting
the central star increases as the semi--major axis decreases.
Nonetheless, in some cases, the interactions (secular or not) with the
outer planets increase the eccentricity of the PAID up to very high
values without the object colliding with the star.  The orbit can then
be tidally circularized by the central star, which leads to a 'hot
PAID'. Six 'hot Saturns' have been observed so far, with $ 0.24 M_{\rm
J} < m \sin i < 0.52 M_{\rm J}$ and $ 0.038 {\rm au} < a < 0.066{\rm
au}.$ Although such a process is expected to be rare within the
context of this model, observational bias against observing low mass
planets at larger distances makes it difficult to assess to what
extent an an alternative mechanism such as direct migration is
required.

If the scenario we have studied in this Letter has opperated, there
should be a massive planet with $a$ on the order of tens of au and a
high eccentricity associated with the eccentric PAID on a close orbit.
In the case represented in figure~\ref{fig5} for instance, the
amplitude of the stellar radial velocity induced by the outer planet
is about 60~m~s$^{-1}$ (that of the inner planet is between 10 and
20~m~s$^{-1}$).  Over a short period of time ($\sim 1$~y), only the
trajectory of the inner planet would be resolved, with the mean value
of the stellar radial velocity drifting by an amount which depends on
whether the outer companion is near apocentre or not, and which varies
between a few m~s$^{-1}$ and tens of m~s$^{-1}$.  Twelve systems have
been selected by Fischer et al. (2001) as candidates for which such
residual velocity drifts are potentially observable, and their results
indicate that the kind of systems we have been describing here are on
the verge of detectability.  Among the systems selected by Fischer et
al. (2001) with a residual drift of 50~m~s$^{-1}$ over a two year
period, HD~38529 has a planet with $m \sin i= 0.76$~M$_{\rm J}$,
$a=0.13$~au and $e=0.27.$ This is similar to the system illustrated in
figure~\ref{fig4}.

\section*{acknowledgment}

J.C.B.P. acknowledges visitor support from the CNRS through a {\em
Poste Rouge} and the IAP for hospitality.



\begin{thebibliography}{}

\bibitem[2000]{Boss1}
Boss, A. P., 2000, ApJ, 536, L101

\bibitem[2001]{Fischer}
Fischer, D. A., Marcy, G. W., Butler, R. P., Vogt, S. S., Frink, S.,
Apps, K., 2001, ApJ, 551, 1107

\bibitem[2001]{Haisch} 
Haisch K. E., Jr., Lada E. A., \& Lada C. J., 2001, ApJ, 553, L153

\bibitem[2000]{Lin} Lin, D. N. C., Papaloizou, J. C. B., Terquem, C.,
Bryden, G., Ida, S., 2000, in Protostars and Planets IV, eds
V. Mannings, A. P. Boss, S. S. Russell (Tucson: Univ. Arizona Press),
p.~1111

\bibitem[1999]{Minu}
Masunaga, H., Inutsuka, S., 1999, ApJ, 510, 822

\bibitem[1980]{Mizuno}
Mizuno, H., 1980, Prog. Theor. Phys., 64, 544

\bibitem[1999]{Murray}
Murray, C. D., Dermott, S. F., 1999, Solar System Dynamics (CUP), p.~254--255

\bibitem[2000]{Nelson}
Nelson, R. P., Papaloizou, J. C. B., Masset, F. S., Kley, W., 2000,
MNRAS, 318,18

\bibitem[2001]{Papaloizou}
Papaloizou, J.C.B., Terquem, C., 2001, MNRAS, 325, 221

\bibitem[1993]{Press}
Press, W.~H., Teukolsky, S.~A., Vetterling, W.~T., Flannery, B.~P., 1993,
Numerical Recipes in FORTRAN
(CUP)

\bibitem[1978]{Roy}
Roy, A. E., 1978, Orbital Motion (Adam Hilger, Bristol), p.~280


\end{thebibliography}
\end{document}